\documentclass[reprint,showpacs,epsf,aps,prb]{revtex4-1}
\usepackage{color,graphicx}
\usepackage{dcolumn}
\usepackage{bm}
\begin{document}
\draft

\title{Anisotropic magnetic properties and giant magnetocaloric effect in antiferromagnetic $R$MnO$_3$ crystals ($R$$=$Dy, Tb, Ho and Yb)}
\author{A. Midya, S. N. Das, and P. Mandal}
\affiliation{Saha Institute of Nuclear Physics, 1/AF Bidhannagar, Calcutta 700 064, India}
\author{S. Pandya and V. Ganesan}
\affiliation{UGC-DAE Consortium for Scientific Research, Khandwa Road,
Indore 452 001, India}

\begin{abstract}
We have systematically investigated the magnetic properties and magnetocaloric effect (MCE) in $R$MnO$_3$ ($R$$=$Dy, Tb, Ho and Yb) single crystals. Above a critical value of applied field ($H_c$), $R$MnO$_3$ undergo a first-order  antiferromagnetic (AFM) to ferromagnetic (FM) transition below the ordering temperature ($T_{N}^{R}$) of $R^{3+}$ moment and a  second-order  FM to paramagnetic (PM)  transition  above $T_{N}^{R}$. Both $H$ and $T$ dependence of $M$ shows that the system is highly anisotropic in the FM as well as PM states and, as a result, the magnetic entropy change ($\Delta S_{M}$) is extremely sensitive to the direction of applied field and can be negative (normal MCE) or positive (inverse MCE). For hexagonal HoMnO$_3$ and YbMnO$_3$ systems, a very small inverse MCE is observed only for $H$ parallel to $c$ axis and it decreases with increasing $H$ and crosses over to normal one above $H_c$.  On the other hand, for orthorhombic DyMnO$_3$ and TbMnO$_3$, though the inverse MCE disappears above $H_c$ along easy-axis of magnetization,  it increases rapidly with $H$ along hard-axis of magnetization for $T$$\ll$$T_{N}^{R}$. Except for YbMnO$_3$, the values of $\Delta S_{M}$, relative cooling power and  adiabatic temperature change along easy-axis of magnetization are quite large in the field-induced FM state for a moderate field  strength. The large values of these parameters, together with negligible hysteresis, suggest that the multiferroic manganites could be potential materials for magnetic refrigeration in the low-temperature region. \\

\end{abstract}

\vskip 1cm
\maketitle

\newpage

\section{Introduction}
In twenty-first century, energy efficient and environmentally friendly technology has received special attention in order to combat the global warming phenomenon and energy crisis. Refrigeration based on the magnetocaloric effect (MCE) has attracted much research interest because of its higher energy efficiency over the conventional vapor compression refrigeration and it does not use ozone-depleting chlorofluorocarbon as a refrigerant \cite{kag}. Magnetocaloric effect  describes the reversible change in temperature  of a  material under adiabatic condition produced by the magnetic entropy change $\Delta$$S_{M}$ due to the variation in applied magnetic field \cite{kag,tishin}. The main aim in this field is to search for new materials, which exhibit a large MCE and are capable of operating at different temperature ranges, depending on the intended applications. Large MCE close to room temperature would be useful for domestic and several technological applications while large MCE in the low-temperature region is important for specific technological applications such as space science and liquefaction of hydrogen in fuel industry \cite{kag,prov}. The guidelines for the choice of an appropriate material are that it should have low heat capacity and exhibit a large entropy change at the ferromagnetic (FM) to paramagnetic (PM) transition or  field-induced metamagnetic transition from antiferromagnetic (AFM) to FM states with a minimal hysteresis.

In colossal magnetoresistive oxides, MCE  has been extensively studied due to their wide variation in the Curie temperature ($T_C$) and nature of FM-PM phase transition \cite{phan,guo,sar}.  On the other hand, MCE in multiferroic manganites $R$MnO$_3$ ($R$$=$Tb to Yb) has not been investigated in details \cite{fab,midya,jin}. So far, most of the studies on these systems are concentrated on the magnetic and ferroelectric properties of Mn sublattice because their interplay has opened up a new dimension in basic research as well as technological application. The magnetic structure determination is a prerequisite for understanding the coupling between ferroelectricity and magnetism in these materials. Several neutron and resonance x-ray studies reveal that $R$MnO$_3$ undergo sequence of complicated magnetic phase transitions with decreasing temperature \cite{que,kimu,kimu1,fabre,nandi,stremp,feyer}. It has also been shown the rare-earth magnetic ordering plays a very important role in the magnetoelectric coupling \cite{yen}. According to the crystallographic structure, these multiferroic  materials are divided into two classes. The compounds with larger rare-earth ion ($R$$=$Tb,Dy) crystallize in orthorhombic structure whereas hexagonal structure is more stable for smaller ionic radius of $R$ ($R$$=$Ho to Lu,Y). In orthorhombic compounds, where the magnetic frustration of Mn spin arises from competing exchange interactions, ferroelectric and magnetic orderings appear at the same temperature \cite{fabre}. However, in hexagonal compounds, the magnetic frustration of Mn spin arises from the lattice geometry since the triangular lattice is frustrated for AFM first-nearest-neighbor interaction and  the ferroelectric order occurs at an elevated temperature ($\sim$900 K) which is well above the magnetic ordering temperature ($\sim$80 K) \cite{fabre}. Irrespective of the magnetic and ferroelectric properties of Mn sublattice,  the AFM ordering of rare-earth moments in multiferroic $R$MnO$_{3}$ occurs at a relatively low temperature ($T_{N}^{R}$$<$8 K) and the magnetic structure is highly anisotropic with very weak  interaction along one of the crystallographic axes. As a result, several multiferroic manganites with high total angular momentum quantum number of rare-earth ion exhibit huge increase in magnetization  at a moderate field strength ($\sim$2 T) which is close to the expected moment  of $R^{3+}$ ion. In this work, we present the field and temperature dependence of magnetic properties of rare-earth sublattice in $R$MnO$_3$ ($R$$=$Dy, Tb, Ho and Yb) crystals.  We observe that these materials (except $R$$=$Yb) exhibit giant MCE and large adiabatic temperature change and RCP  due to the field-induced AFM-FM transition.  This opens up a possibility for another viable technological application for multiferroic manganites namely, in magnetic cooling at low temperature.  Furthermore, we have shown  that one can clearly differentiate orthorhombic manganites from hexagonal ones because of their distinct magnetic and magnetocaloric properties related to rare-earth ion.
\section{Experimental techniques and sample preparation}
Polycrystalline $R$MnO$_{3}$ ($R$$=$Tb, Dy, Ho and Yb) samples were prepared from stoichiometric mixture of $R$$_{2}$O$_{3}$ and Mn$_{3}$O$_{4}$ by solid-state reaction and the single crystals were grown from the  polycrystalline rod by travelling solvent floating zone technique using an image furnace (NEC) \cite{pm,midya}. Magnetic and heat capacity measurements were carried out employing  superconducting quantum interference device magnetometer (Quantum Design) and physical properties measurement system (Quantum Design). The x-ray diffraction patterns of powdered sample of single crystals reveal that these materials are single phase. The Rietveld profile refinement of diffraction patterns shows that DyMnO$_{3}$ and TbMnO$_{3}$ have orthorhombic Pbnm structure whereas HoMnO$_3$ and YbMnO$_{3}$ exhibit hexagonal crystal structure with space group P6$_{3}$cm. Several magnetic parameters such as magnetization ($M$), susceptibility ($\chi$) and entropy ($S_M$) with field parallel to easy axis ($H$$\Vert$$e$) are denoted as $M_e$, $\chi_e$ and $S_{Me}$, respectively whereas the respective parameters with field parallel to hard axis ($H$$\Vert$$h$) are denoted as $M_h$, $\chi_h$ and $S_{Mh}$ for both orthorhombic and hexagonal structures.
\section{Results and Discussions}
\begin{figure}
  \includegraphics[width=0.4\textwidth]{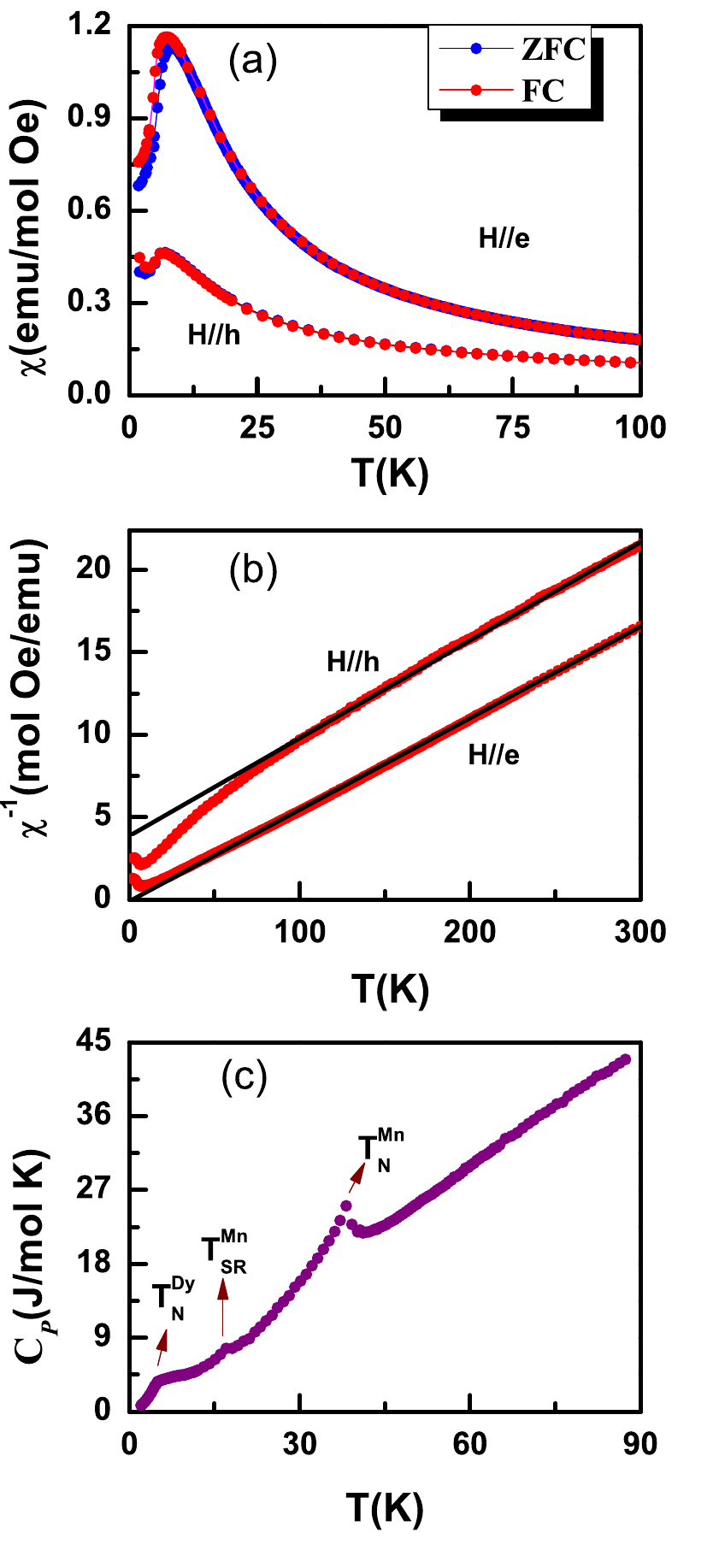}
  \caption{(a) Temperature dependence of dc susceptibility along the easy-axis ($b$-axis) and hard-axis ($a$-axis) of DyMnO$_3$. (b) Temperature dependence of $\chi$$^{-1}$ along both axes. Solid lines are the linear fit to the data along the respective axis. (c) Thermal evolution of heat capacity of DyMnO$_3$ crystal at zero field.\\}\label{Fig.1}
\end{figure}
The isothermal magnetic entropy change $\Delta S_{M}$ can be calculated from $H$ and $T$ dependence of $M$ using the Maxwell relation, $\Delta S_{M}$($T,H$)$=$$\int\limits_{0}^{H} \left(\frac{\partial M}{\partial T}\right)_H dH$ \cite{kag,tishin}. Since the magnetization measurements are performed at discrete intervals of $T$ and $H$, $\Delta S_{M}$ is numerically calculated approximately using the expression, $\Delta S_{M}$($T,H$) $=$ $\sum_{i}\frac{M_{i+1} - M_{i}}{T_{i+1} - T_{i}}$$\Delta$H$_{i}$, where M$_{i+1}$ and M$_i$ are the experimental values of magnetization measured with a field H$_{i}$ at temperatures T$_{i+1}$ and T$_{i}$, respectively. The characteristic parameter which determines the magnetic cooling efficiency of a magnetocaloric material is the relative cooling power (RCP) and is defined as,
\begin{equation}
RCP=\int\limits_{T_1}^{T_2}\Delta S_{M}dT\
\end{equation}
where T$_1$ and T$_2$ are the temperatures corresponding to both sides of the half-maximum value of $\Delta S_{M}$ peak. The relative cooling power is the measure of the amount of heat transfer between the cold and hot reservoirs in an ideal refrigerator as a function of field. The adiabatic temperature change ($\Delta$$T_{ad}$),  another important parameter related to MCE, can be calculated from the field-dependent magnetization and zero-field heat capacity data. The total entropy $S$(0,$T$) in absence of magnetic field is given by
\begin{equation}
S(0, T)=\int\limits_{0}^{T}\frac{C(0, T)}{T}dT\
\end{equation}
and then $S(H, T)$ may be evaluated by subtracting the corresponding $\Delta S_{M}$ from $S(0, T)$. The isentropic temperature change between the entropy curves $S(0, T)$ and $S(H, T)$ provides the value of  $\Delta T_{ad}$($T$) \cite{vkp}.
\subsection{Orthorhombic DyMnO$_3$ and TbMnO$_3$ systems}
\begin{figure}
  \includegraphics[width=0.4\textwidth]{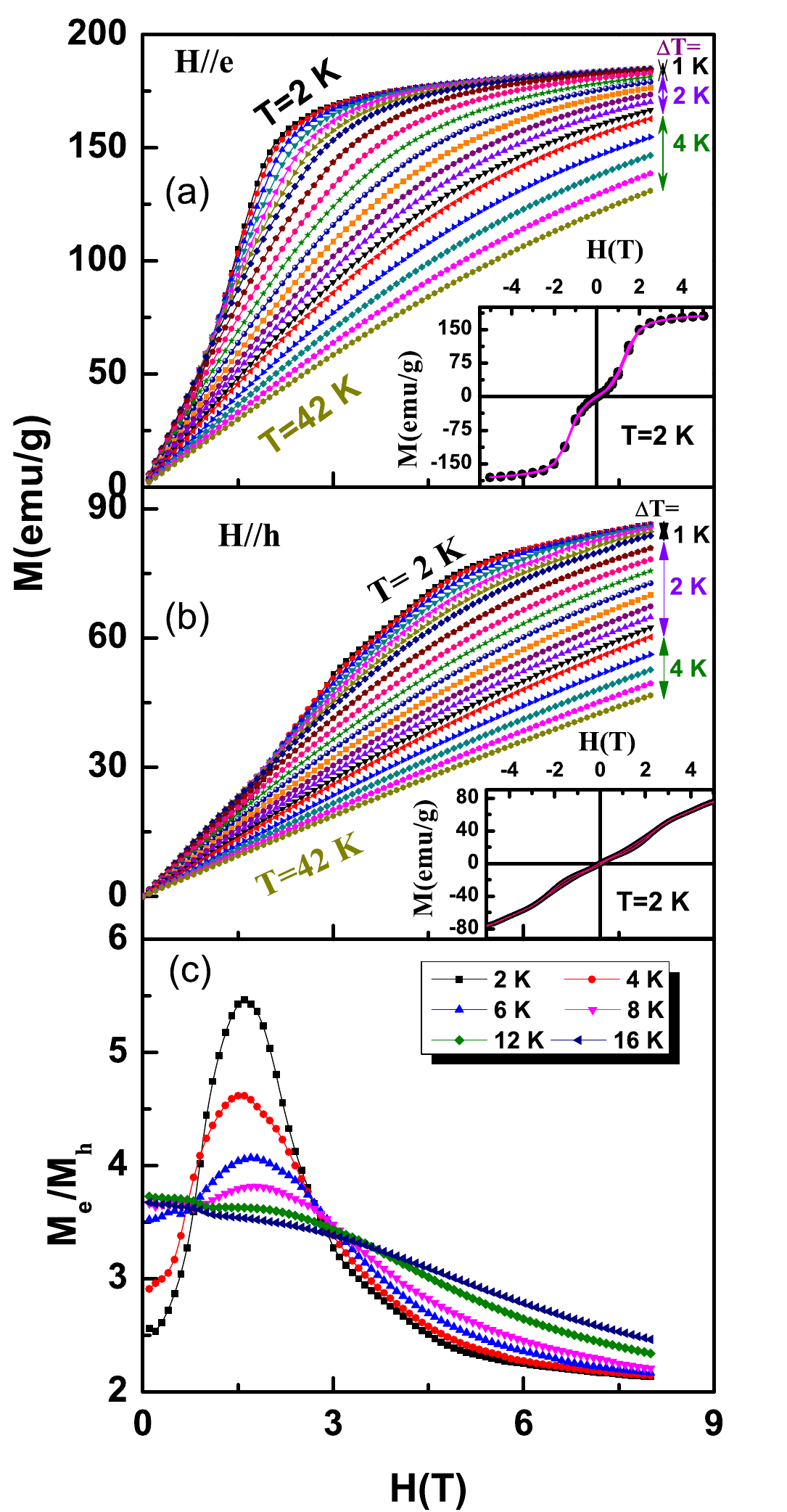}
  \caption{Isothermal magnetization of DyMnO$_3$ as a function of magnetic field for different temperatures  along the easy (a) and  hard axes (b). Insets show the hysteresis at 2 K. Ratio of isothermal magnetization along the easy and hard axes ($M_e/M_h$) as a function of $H$ at some selective temperatures (c).}\label{Fig.2}
\end{figure}
The temperature dependence of zero-field-cooled (ZFC) and field-cooled (FC) dc susceptibility $\chi$ ($=$$M$/$H$) for $H$ ($=$10 Oe) along the easy-axis  ($b$ axis) and hard-axis ($a$ axis) of magnetization  are shown in Fig. 1(a). It is clear from  the figure  that  DyMnO$_{3}$ is magnetically anisotropic and undergoes a PM to AFM transition below $T_{N}^{Dy}$$\sim$6.5 K  due to the long-range ordering of the Dy$^{3+}$ moments.  For further understanding the nature of magnetic interaction, we have plotted $\chi^{-1}$ versus $T$ [Fig. 1(b)].  The linearity of $\chi^{-1}$($T$) over a wide temperature range suggests that susceptibility follows the Curie-Weiss (CW) law [$\chi$$=$C/($T-\theta_{cw}$)]. From the high temperature linear fit, we have calculated the CW temperature $\theta_{cw}$$\sim$2 K and the effective moment $P_{eff}$$=$11.8 $\mu$$_{B}$ for $H$$\Vert$$e$ while the corresponding values are $-$66 K and 11.6 $\mu$$_{B}$ for $H$$\Vert$$h$.  The observed values of $P_{eff}$ are close to the theoretically expected moment 11.7 $\mu$$_{B}$, calculated using the two-sublattice model $P_{eff}$=$\sqrt{(P^{Dy}_{eff})^{2}+(P^{Mn}_{eff})^{2}}$. The large difference in $\theta_{cw}$ for two different crystallographic axes reflects the anisotropic nature of the exchange interaction.  The small value of $\theta_{cw}$ indicates that the  magnetic interaction along the easy axis is very weak. On the other hand, the large and negative value of $\theta_{cw}$ indicate that the magnetic interaction along the hard axis is strong and antiferromagnetic in nature. For $T$$\gg$$T_{N}^{Dy}$, the large difference in $\chi$($T$) between two axes  implies that the system remains highly anisotropic even in the PM state.  Fig. 1(c) shows the temperature dependence of specific heat in zero field. The anomalies in  $C_p$  at 38.2, 17.1  and 5.5 K correspond to  AFM transition of Mn moment ($T_{N}^{Mn}$) into a sinusoidal incommensurate phase, lock-in transition ($T_{SR}^{Mn}$) of Mn spin and AFM transition of Dy$^{3+}$ ion, respectively.\\
\begin{figure}
  \includegraphics[width=0.5\textwidth]{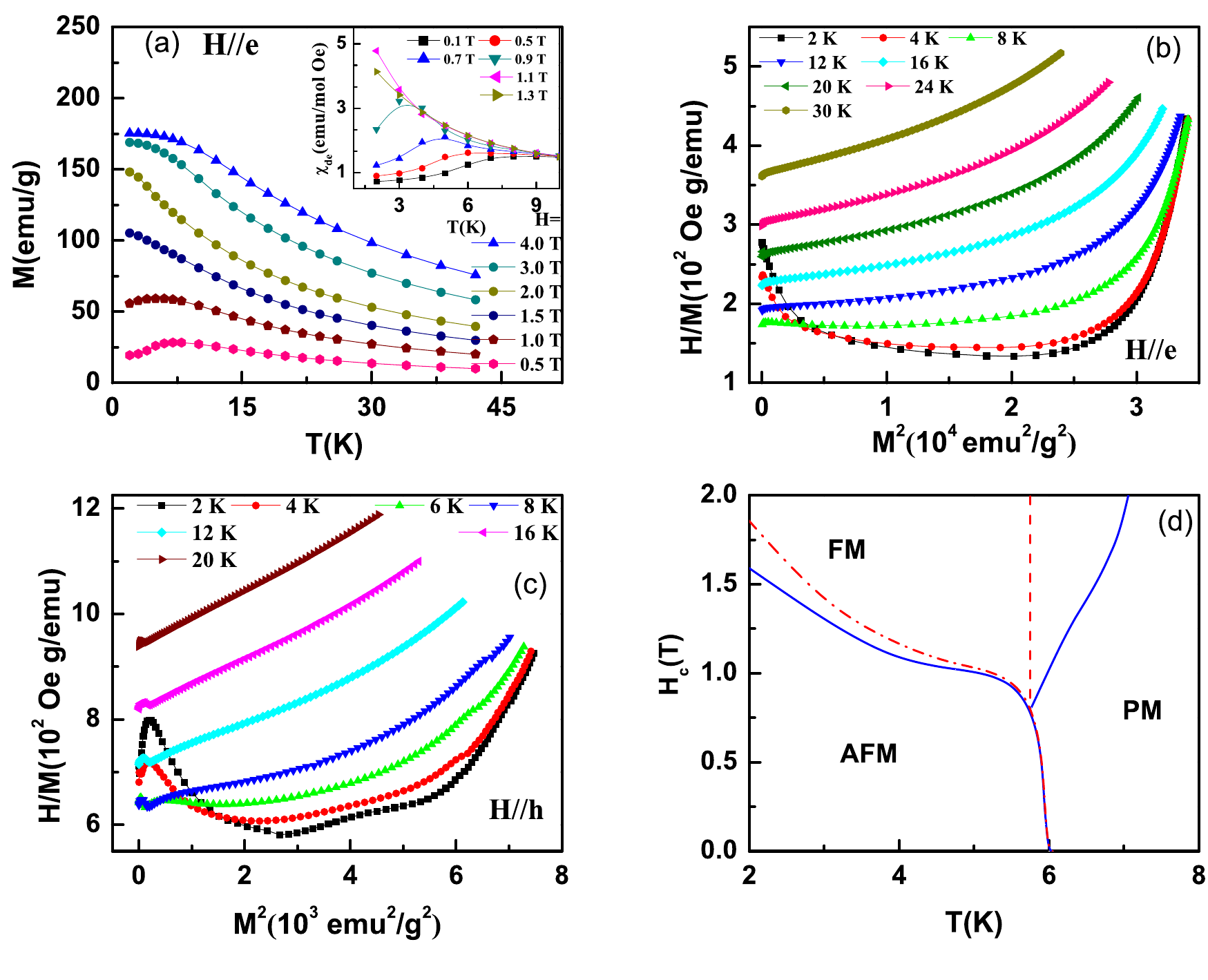}
  \caption{(a) Temperature dependence of magnetization along the easy-axis of DyMnO$_3$ for different magnetic fields. Inset shows the differential susceptibility ($\chi$$_{de}$) versus temperature for different fields. (b) The Arrott plots with $H$ along the easy axis. (c) The Arrott plots with $H$ along the hard axis. (d) The $H-T$ phase diagram of DyMnO$_3$.}\label{Fig.3}
\end{figure}
For elucidating  the role of applied magnetic field on AFM ordering, we have measured $H$ dependence of $M$ in the vicinity of $T_{N}^{Dy}$ and beyond. The field is applied nearly parallel to the easy- and hard-axis of magnetization.  For each isotherm, the magnetic field has been varied from 0 to 8 T. Some representative plots of isothermal field variation of $M$ in the temperature range 2-42 K are shown in Figs. 2(a) and (b), which depict a field-induced metamagnetic transition.   From these plots, one can clearly differentiate the nature of $H$ dependence of $M$ below $T_{N}^{Dy}$ from that above $T_{N}^{Dy}$. Below $T_{N}^{Dy}$,  for both the axes, $M$ increases slowly with $H$ in the low-field region followed by a sharp jump at a critical field $H_c$  and then increases slowly with further increase of $H$. $M$ does not show  monotonic temperature dependence for $H$$<$$H_c$. This behavior is consistent with the field-induced transition from AFM to FM state at $H$$=$$H_c$. We observe that $H_c$  is slightly higher and the field-induced metamagnetic transition is less sharper for $H$$\Vert$$h$ as compared to $H$$\Vert$$e$. Also, the value  and  nature of $H$ dependence of $M$ for $H$$\Vert$$h$ and $H$$\Vert$$e$ are significantly different. This is more clearly reflected in the field dependence of the ratio $M_e/M_h$, which shows a sharp and symmetric peak at around 1.5 T [Fig. 2(c)]. As $T$ increases, the peak becomes broad and shifts slowly towards higher $H$. Above 10 K, the peak changes into a shoulder-like feature. Another important point to be mentioned here is that $M_h$($H$) curve shows a weak feature at around 5 T  below $T_{N}^{Dy}$; indicating the presence of a second transition. Similar weak feature at high field has also been reported earlier in single crystals of DyMnO$_3$ \cite{stremp}.  The insets of Figs. 2(a) and (b) display the five-segment $M$($H$) loop at 2 K. We did not observe any hysteresis at low field.  This is expected because DyMnO$_3$ is AFM  for $H$$<$1.5 T. However, polycrystalline samples with small grain size may show a weak hysteresis due to the surface ferromagnetism \cite{gals}. As the size of crystallites is quite large in good quality single crystals, the contribution from surface magnetism is very small. The value of saturation magnetization ($M_s$) deduced at 2 K and 8 T is  8.8 $\mu_B$ for  $H$$\Vert$$e$ which is  88$\%$ of expected moment (10 $\mu_B$). In contrary to this, $M$ does not show saturation-like behavior and  its value is less than half of the expected moment for $H$$\Vert$$h$. The  higher value of $H_c$, low magnetic moment and  absence of  saturation-like behavior are consistent with the large and negative value of  $\theta_{cw}$ for $H$$\Vert$$h$.
\begin{figure}
  \includegraphics[width=0.35\textwidth]{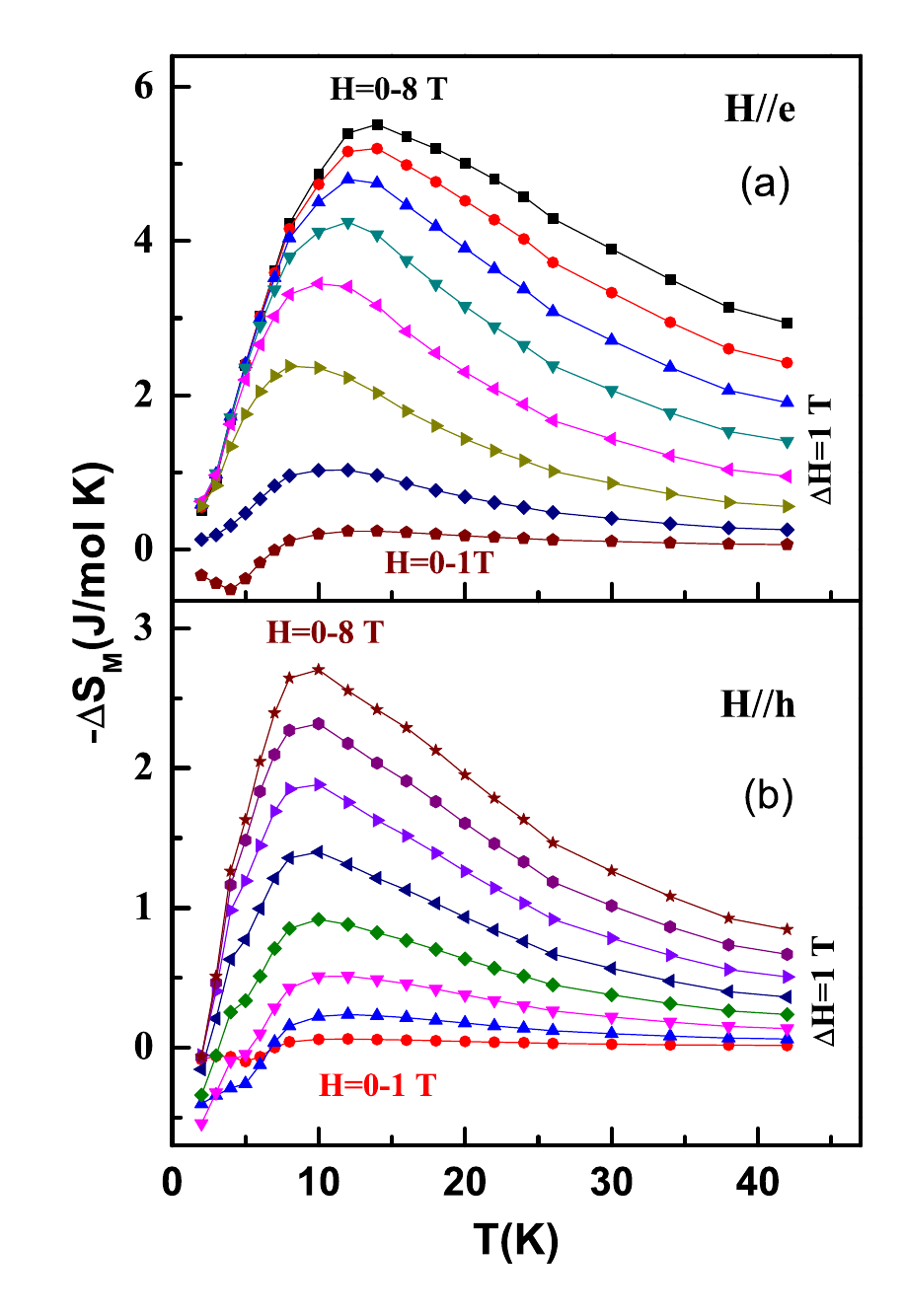}
  \caption{Magnetic entropy change versus temperature for different field change along (a) easy axis (b) hard axis.}\label{Fig.4}
\end{figure}
The temperature dependence of $M$ and differential susceptibility $\chi_d$ ($=$d$M$/d$H$) are useful to understand the field-induced magnetic transition. Figure 3(a) shows $T$ dependence of $M$ and $\chi_{d}$ for $H$$\Vert$$e$.  The peak position in $M_e$($T$) and $\chi_{de}$($T$) curves corresponds to AFM transition temperature $T_{N}^{Dy}$. $T_{N}^{Dy}$ decreases rapidly with increasing $H$ and disappears at  $H_c$ in both the cases.  Above $H_c$, the nature of $M$($T$) curve changes dramatically. $M$ increases rapidly with decreasing $T$ but  saturates at low temperatures. The saturation region  widens  with the increase of field strength. These behavior suggest that  DyMnO$_{3}$ undergoes field-induced AFM-FM transition below $T_{N}^{Dy}$ and PM-FM transition above $T_{N}^{Dy}$. Normally, the field-induced order-order transition is first order in nature.  For understanding the nature of AFM-FM and  FM-PM phase transitions, we have transformed the $M_e$($H$) data into Arrott plots as shown in Fig. 3(b) \cite{arrot}. The slope of $H/M$ versus $M^2$ curve is useful to determine the order of both temperature  and field driven magnetic phase transition. The negative slope of the Arrott plot often indicates a first-order nature of the transition, while the  positive slope implies a second-order transition \cite{sk}. In the present case, the negative slope below $T_{N}^{Dy}$ for $H$$<$1.8 T indicates  that the field-induced AFM-FM transition is first-order in nature while the positive slope of the high-field data  and their linear extrapolation  to $H$$=$0 at a non-zero positive value of $M$ for $T_{N}^{Dy}$$\leq$$T$$\leq$20 K suggest the second-order nature of  the PM-FM transition.  The Arrott plots are also done for $M_h$($H$) data [Fig. 3(c)]. Though the Arrott plots for $H$ along two different directions are qualitatively similar below $T_{N}^{Dy}$, the nature of high-field data above $T_{N}^{Dy}$ in two cases are very different.  The first-order nature of AFM-FM transition is also evident from the low-field $M_h$ data. However, the linear extrapolation of high-field $M_h$ data to $H$$=$0 does not reveal any non-zero and positive value of $M$ for $T$$>$$T_{N}^{Dy}$. This suggests that the PM-FM phase boundary at $T_{N}^{Dy}$ is extremely sharp when the field is applied along the hard-axis of magnetization. The temperature dependence of $H_c$, determined from the maximum in d$M$($H$)/d$H$ (for $T$$<$$T_{N}^{Dy}$) and minimum in d$M$($T$)/d$T$ (for $T$$>$$T_{N}^{Dy}$) curves and the Arrott plots, is summarized in the ($H-T$) phase diagram [Fig. 3(d)].
\begin{figure}
  \includegraphics[width=0.35\textwidth]{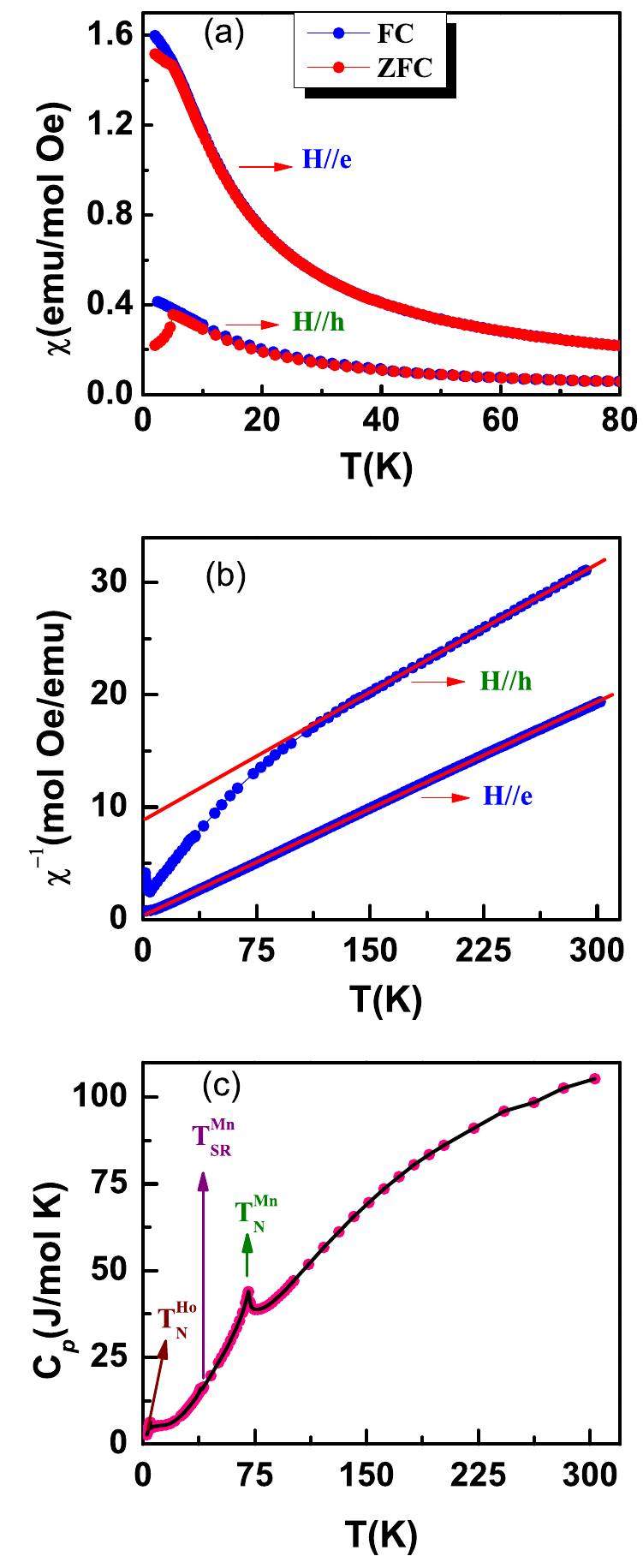}
  \caption{(a) Temperature dependence of FC and ZFC dc susceptibility curves for HoMnO$_3$ along easy axis ($a$ axis) and hard axis ($c$ axis). (b) Inverse susceptibility versus temperature measured along easy and hard axis. Solid line indicates the Curie-Weiss fit in both directions. (c) Temperature dependence of specific heat for HoMnO$_3$ crystal at zero magnetic field.}\label{Fig.5}
\end{figure}
In order to test whether DyMnO$_{3}$ is a suitable candidate for magnetic refrigeration, we have calculated the isothermal magnetic entropy change from the  $M$($H$) curves [Figs. 2(a) and (b)] using the Maxwell relation. Figure 4 presents the thermal distribution of  $\Delta S_{M}$ for field variation up to 8 T for  $H$ parallel to easy and hard axes. In both the cases, $\Delta S_{M}$ is negative above $T_{N}^{Dy}$   and the magnitude of -$\Delta S_{M}$($T$) at maximum ($\Delta S_{M}^{max}$)  increases with  field. $\Delta S_{M}^{max}$ is as high as   5.52 J mol$^{-1}$ K$^{-1}$ at 8 T for $H$$\Vert$$e$. However, the $H$ dependence of $\Delta S_{M}$ below $T_{N}^{Dy}$ are quite different in two cases.  $\Delta S_{Me}$ is positive (inverse MCE) below $T_{N}^{Dy}$ only for small field changes $\Delta H$$<$$H_c$ ($=$1.5 T). On the other hand, $\Delta S_{Mh}$ remains positive up to 8 T for $T$$<$3 K. Moreover, the steep decrease in $\Delta S_{Mh}$($T$) on the low-temperature side of the maximum suggests that DyMnO$_{3}$ may exhibit a large inverse MCE below 2 K.  It is worth noting that the position of the maximum in -$\Delta S_{M}$($T$) curve shifts slowly towards higher temperature with increasing field for $H$$\Vert$$e$ while it is insensitive to field for $H$$\Vert$$h$. This behavior is consistent with the observed ($H-T$) phase diagram. For $H$$\Vert$$e$, the positive value of  $\Delta S_{Me}$ below $T_{N}^{Dy}$ initially  increases and then decreases with the increase of $H$ and eventually becomes negative above 1.5 T. Such type of $H$ dependence of $\Delta S_{M}$ is quite common in antiferromagnetic systems where the field-induced AFM-FM transition occurs \cite{kag}. The initial increase in $\Delta S_{M}$ with $H$ below $H_c$ is due to the field-induced magnetic disordering. When an external magnetic field is applied along the easy axis, the magnetic moment fluctuation is enhanced in one of the two AFM sublattices which is antiparallel to $H$. With the increase of $H$, more and more spins in the antiparallel sublattice orient along the field direction. This, in turn, increases the spin disordering and it continues up to a certain field below $H_c$. As the system becomes  ferromagnetic for $H$$>$$H_c$, the majority of spins in the antiparallel sublattice orient along the field direction and, as a consequence, $\Delta S_{M}$ becomes negative. The relative cooling power is evaluated to determine the cooling efficiency of DyMnO$_3$ crystal as a magnetocaloric material.  RCP is quite large (155 J mol$^{-1}$) for a field change of 8 T for $H$$\Vert$$e$. We have also estimated the adiabatic temperature difference from the isentropic curves using the zero-field specific heat data in Eq. (2).  $\Delta$$T_{ad}$ is found to be as high as 11.5 K for a field change of 8 T for $H$$\Vert$$e$. This value of $\Delta$$T_{ad}$ is appreciably larger than that observed in perovskite manganites \cite{kag,phan}.
\begin{figure}
  \includegraphics[width=0.5\textwidth]{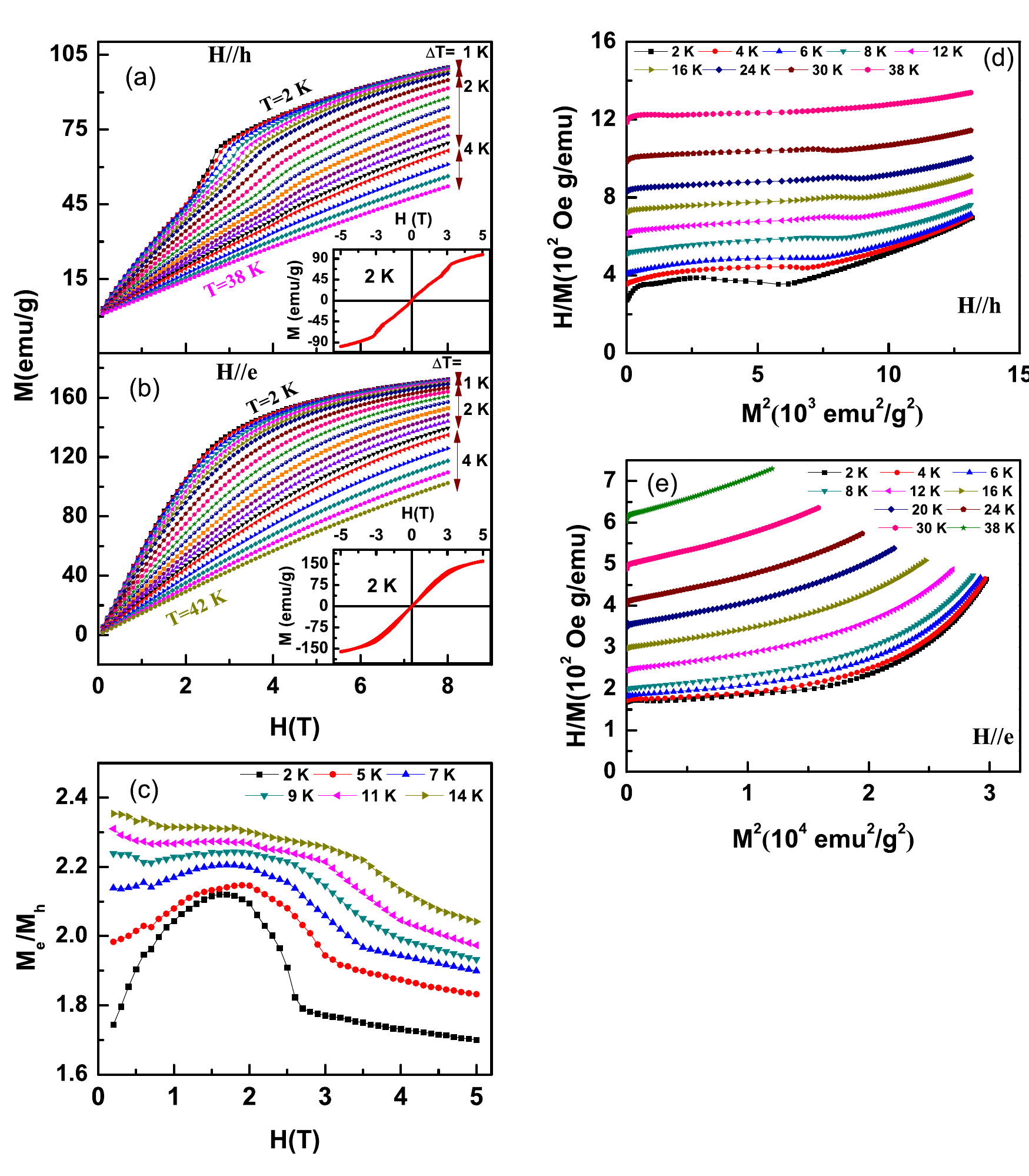}
  \caption{Figure 6. Field dependence of isothermal magnetization for HoMnO$_3$ with $H$ parallel to hard axis (a) and  easy axis (b). $M_e/M_h$ versus $H$ plot at some selective temperatures (c). The Arrott plots of the magnetization isotherms along  hard axis (d) and easy axis (e).}\label{Fig.6}
\end{figure}
Similar to DyMnO$_{3}$, the temperature and field dependence of magnetization with $H$ along easy-axis ($a$ axis) and hard-axis ($b$ axis) of TbMnO$_{3}$ crystal has been measured. We observe that the $T$  and $H$  dependence of magnetization for TbMnO$_{3}$ crystal is similar to earlier reports \cite{jin,que,kimu}. Unlike DyMnO$_{3}$, the inverse MCE  in TbMnO$_{3}$ is appreciably large and increases with $H$ along $b$ axis. This suggests that the AFM ground state in TbMnO$_{3}$ is quite stable against applied field along the hard axis of magnetization. In spite of this difference, the nature of field-induced AFM-FM and PM-FM transitions and ($H-T$) phase diagrams for $H$$\Vert$$e$ as well as for $H$$\Vert$$h$ are qualitatively similar in two systems. Also,  we observe that the values of magnetocaloric parameters $\Delta S_{M}^{max}$, RCP and $\Delta$$T_{ad}$  in DyMnO$_{3}$ and TbMnO$_{3}$ are comparable in magnitude. Therefore, both DyMnO$_{3}$ and TbMnO$_{3}$ satisfy the major important criteria for  magnetic refrigeration at low temperatures.
\subsection{Hexagonal HoMnO$_3$ and YbMnO$_3$ systems}
\begin{figure}
  \includegraphics[width=0.35\textwidth]{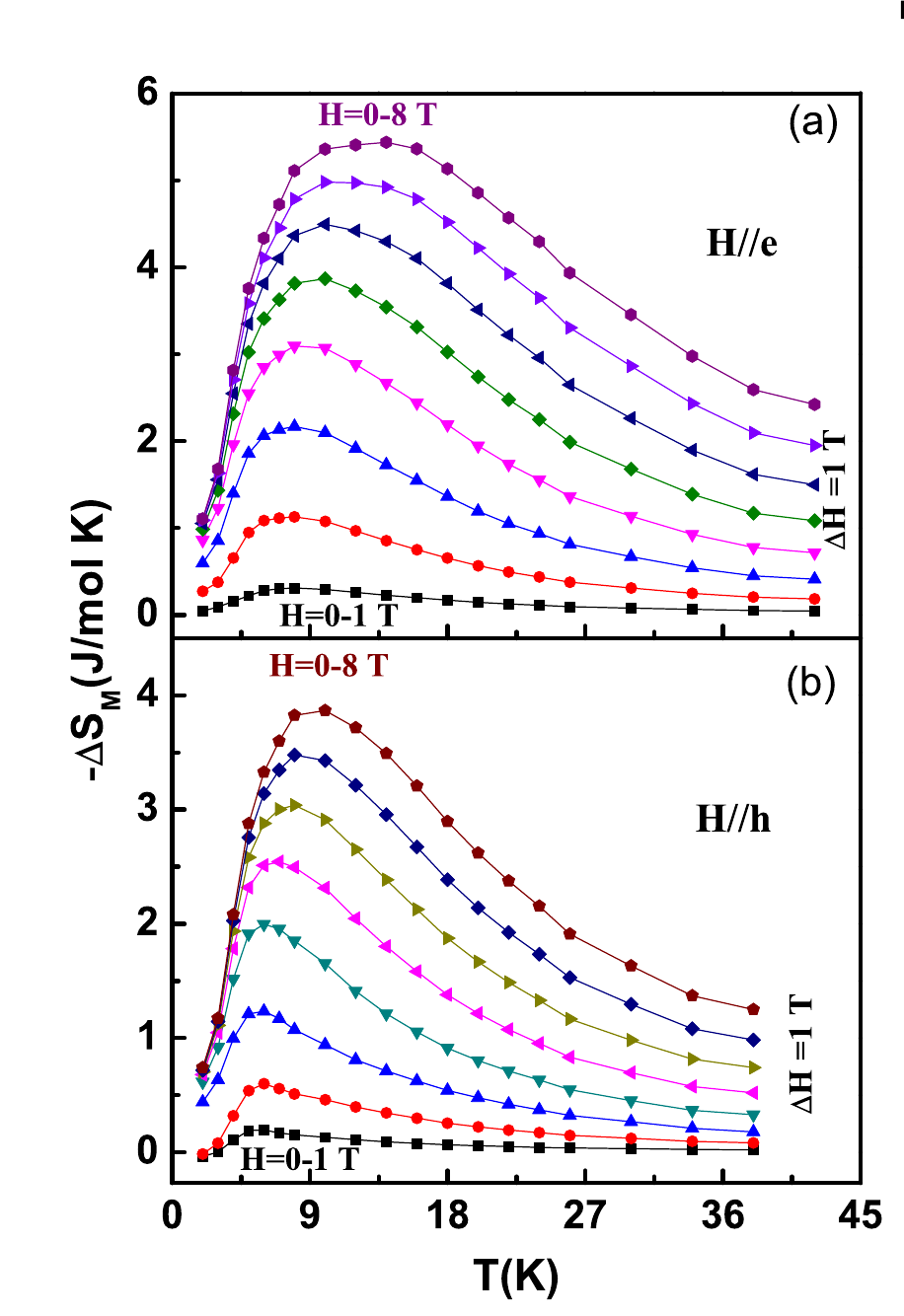}
  \caption{Figure 7. Temperature variation of magnetic entropy change for different field change along (a) easy axis (b) hard axis for HoMnO$_3$.}\label{Fig.7}
\end{figure}
The temperature dependence of low-field ($H$$=$4 Oe) susceptibility  along the two principal axes of hexagonal unit cell of HoMnO$_3$ is shown in Fig. 5(a). The anomaly at 4.5 K in ZFC cycles is due to the AFM ordering of Ho moments whereas the weak bifurcation between FC and ZFC curves below $\sim$40 K corresponds to the reorientation of Mn moment in the basal plane perpendicular to the initial direction.  No anomaly is observed at the AFM ordering temperature of the Mn moments. Figure shows that the values of $\chi$  at any temperature are significantly different along two  crystallographic directions; reflecting the anisotropic nature of the magnetic structure of HoMnO$_3$.  For the quantitative estimation of anisotropy in magnetic parameters, we have presented $\chi^{-1}$($T$) for $H$ along easy axis ($a$ axis) and hard axis ($c$ axis) [Fig. 5(b)]. Both the curves show that $\chi$ follows the CW behavior over a wide range of $T$. From the linear part of $\chi^{-1}$($T$) for $H$$\Vert$$h$, the fitted values of $P_{eff}$ and  $\theta_{cw}$ are found to be 10.24 $\mu_{B}$ and $-$116 K, respectively while the corresponding values are 11.2 $\mu_{B}$ and $-$5 K for $H$$\Vert$$e$ which are consistent with the previously reported results \cite{sugi,loren,loren1,vajk}.  Significant anisotropy in magnetic interaction is evident from the large difference in the values of $P_{eff}$ and $\theta_{cw}$ for two principal axes of the unit cell. Unlike susceptibility, the temperature variation of zero-field heat capacity is showing three distinct transitions [Fig. 5(c)]. The peak at 4.6 K indicates the AFM ordering of Ho moment as evidenced from magnetic measurement. The small peak at $\sim$39 K is due to the Mn moment reorientation and the third $\lambda$-type anomaly originates from the AFM ordering of the Mn$^{3+}$ magnetic moments.
\begin{figure}
  \includegraphics[width=0.35\textwidth]{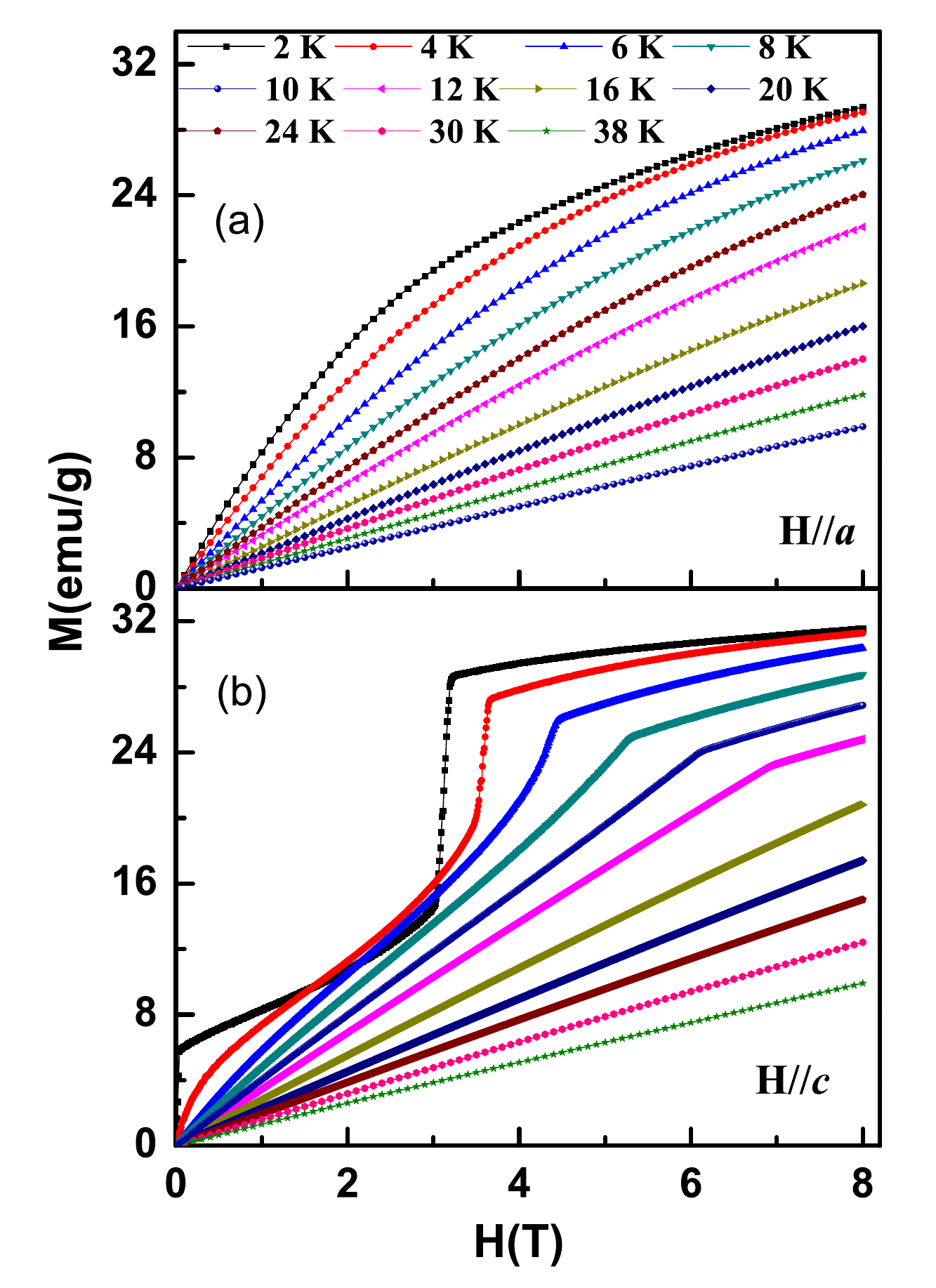}
  \caption{Figure 8. Isothermal magnetization of  YbMnO$_3$ as a function of magnetic field for different temperatures  with $H$ along the $a$ axis (a) and $c$ axis (b).}\label{Fig.8}
\end{figure}
Figures 6(a) and (b) present the  field dependence of isothermal magnetization for HoMnO$_3$ with field parallel to $c$ and $a$  axes, respectively.
We would like to mention that $M$ (at 3 K and 5 T) is 20-25$\%$ larger for the present sample than the reported value on HoMnO$_3$ single crystal grown by the flux method \cite{gals}. The $M$($H$) isotherms for $H$$\Vert$$h$ indicate a field-induced metamagnetic transition. At low fields, the dependence of $M$ on $H$ is approximately linear and the slope of $M$($H$) curve decreases above a critical field but no saturation is attained up to 8 T. To investigate the reversibility of the field-induced magnetic transition, we have measured the five-segment $M$($H$) loop at 2 K and observed no significant hysteresis  [Inset of Fig. 6(a)]. The nature of $M$($H$) curves for $H$$\Vert$$e$ differs from those for $H$$\Vert$$h$. For $H$$\Vert$$e$, $M$($H$) curve  displays no  abrupt change but increases smoothly with $H$ and the value of $M$ is significantly larger as compared to that for $H$$\Vert$$h$. The anisotropy in the field-induced FM state is apparent from  the $M$$_{e}$/$M_{h}$ versus $H$ plot for different $T$ [Fig. 6(c)]. The field and temperature dependence of $M_{e}$/M$_{h}$ for HoMnO$_3$ may be compared with that for orthorhombic DyMnO$_{3}$. One can clearly see that there are several important differences between the two systems. For example, in DyMnO$_3$, the peak in $M_{e}$/M$_{h}$ versus $H$ plot is  very sharp at low temperature. On the other hand, the maximum for HoMnO$_3$ is quite broad and its height is almost half of that for DyMnO$_3$. Also, the nature of $T$ dependence of the peak are quite different in two cases. With increasing $T$, though the peak broadens, the value of $M_{e}$/M$_{h}$ at peak increases slowly in HoMnO$_3$ while the peak decreases and  broadens rapidly  in DyMnO$_3$. Figures 6(d) and (e) show the Arrott plots for field along both the axes.  The Arrott plots  show that the field-induced AFM-FM transition is first order while PM-FM transition is second order and the PM-FM phase boundary is quite sharp for field along hard-axis of magnetization [Fig. 6(d)]. However, closer inspection shows that the nature of the Arrott plots for $H$$\Vert$$e$ [Fig. 6(e)]  differs significantly from that for $H$$\Vert$$h$  and also from that for orthorhombic systems. For $H$$\Vert$$e$, $H/M$ versus $M^2$  curves do not show negative slope in low-field region as expected for the first-order transition. This is due to the smooth increase of $M$ with $H$ [Fig. 6(b)]. For $H$$\Vert$$h$, the $H/M$ versus $M^2$ curves for HoMnO$_3$ are almost linear in the high-field region both above and below $T_{N}^{Ho}$. Normally, the $H/M$ versus $M^2$ curves are linear and parallel for mean-field like FM-PM transition otherwise a upward curvature is exhibited as in the case of $H$$\Vert$$e$. A small deviation from parallelism may occur due to the weak field dependence of $T_C$ in the high-field region. Thus it appears that the system may belong to different universality class depending on the direction of applied magnetic field.
\begin{figure}
  \includegraphics[width=0.35\textwidth]{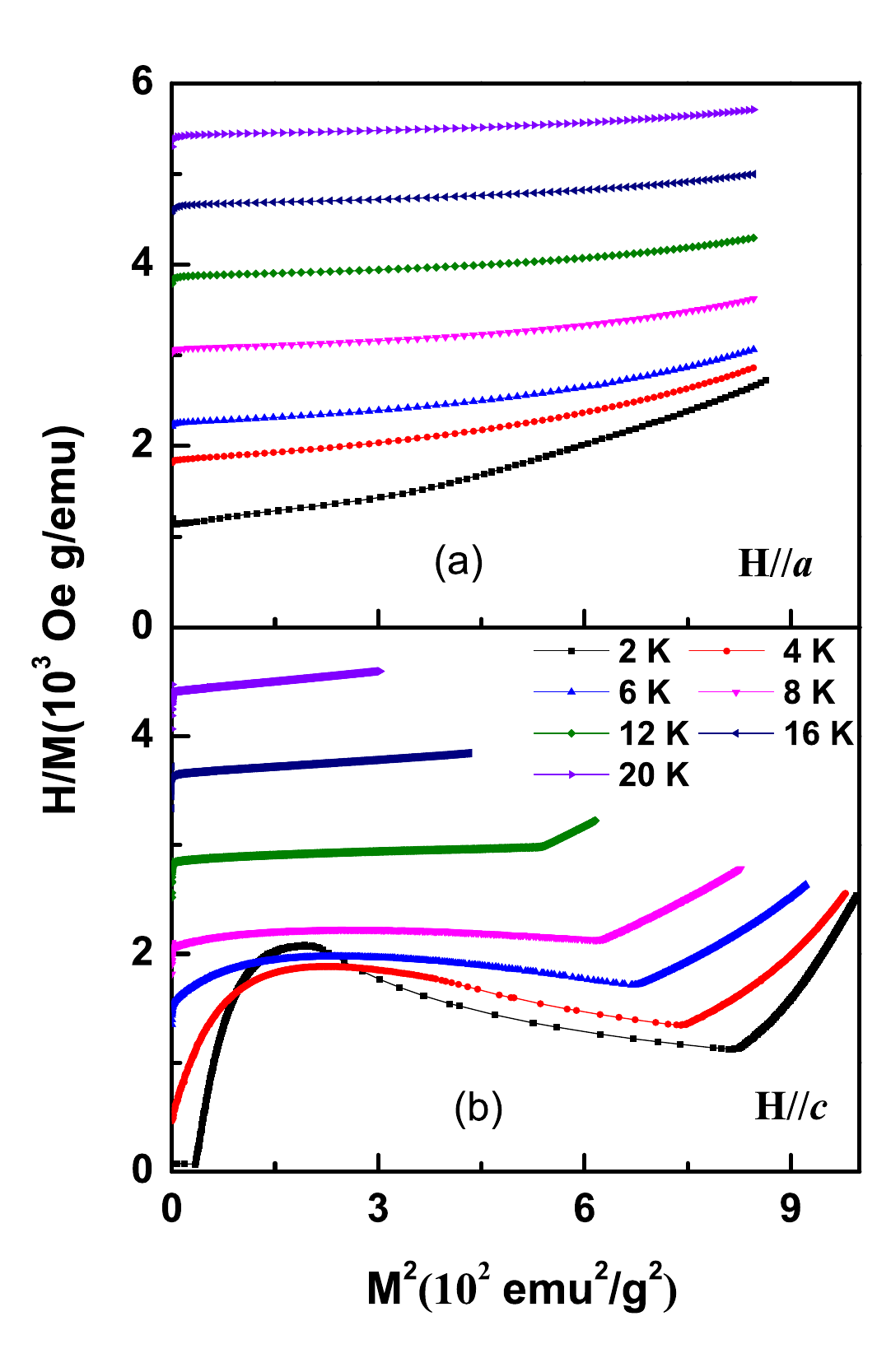}
  \caption{Figure 9. The Arrott plots of  the magnetization isotherms  for YbMnO$_3$ with field direction parallel to $a$ axis (a) and $c$ axis (b).}\label{Fig.9}
\end{figure}
For HoMnO$_3$, the  isothermal entropy change for different field variations was estimated. Figures 7(a) and (b) depict the nature of $\Delta S_{M}$($T$) curves for $H$$\Vert$$e$ and $H$$\Vert$$h$, respectively. In both the cases, a broad maximum is observed. Apart from the values, there are several important differences in the nature of $H$ dependence of $\Delta S_{M}$ for two crystallographic axes. With the increase of magnetic field,  the maximum in $\Delta S_{M}$($T$) shifts to higher temperature for $H$$\Vert$$e$ while the position of the maximum remains pinned at $\sim$5 K  for $H$$\Vert$$h$. For $H$$\Vert$$e$, $\Delta S_{M}$ is negative down to 2 K even at very low fields, i.e., inverse MCE is absent in the AFM state. However, for $H$$\Vert$$h$, $\Delta S_{M}$ is positive well below $T_{N}^{Ho}$ at low fields ($H$$<$$H_c$). Unlike DyMnO$_3$, inverse MCE in HoMnO$_3$ is very small.  One can see that $\Delta$S$_{M}^{max}$ is large for a moderate field change along easy axis.  The values of other magnetocaloric parameters such as RCP  and $\Delta T_{ad}$ are also large for HoMnO$_3$ single crystals. We observe RCP$=$144 J mol$^{-1}$ and $\Delta T_{ad}$$=$12.5 K for the field change of 8 T along easy axis.
\begin{figure}
  \includegraphics[width=0.35\textwidth]{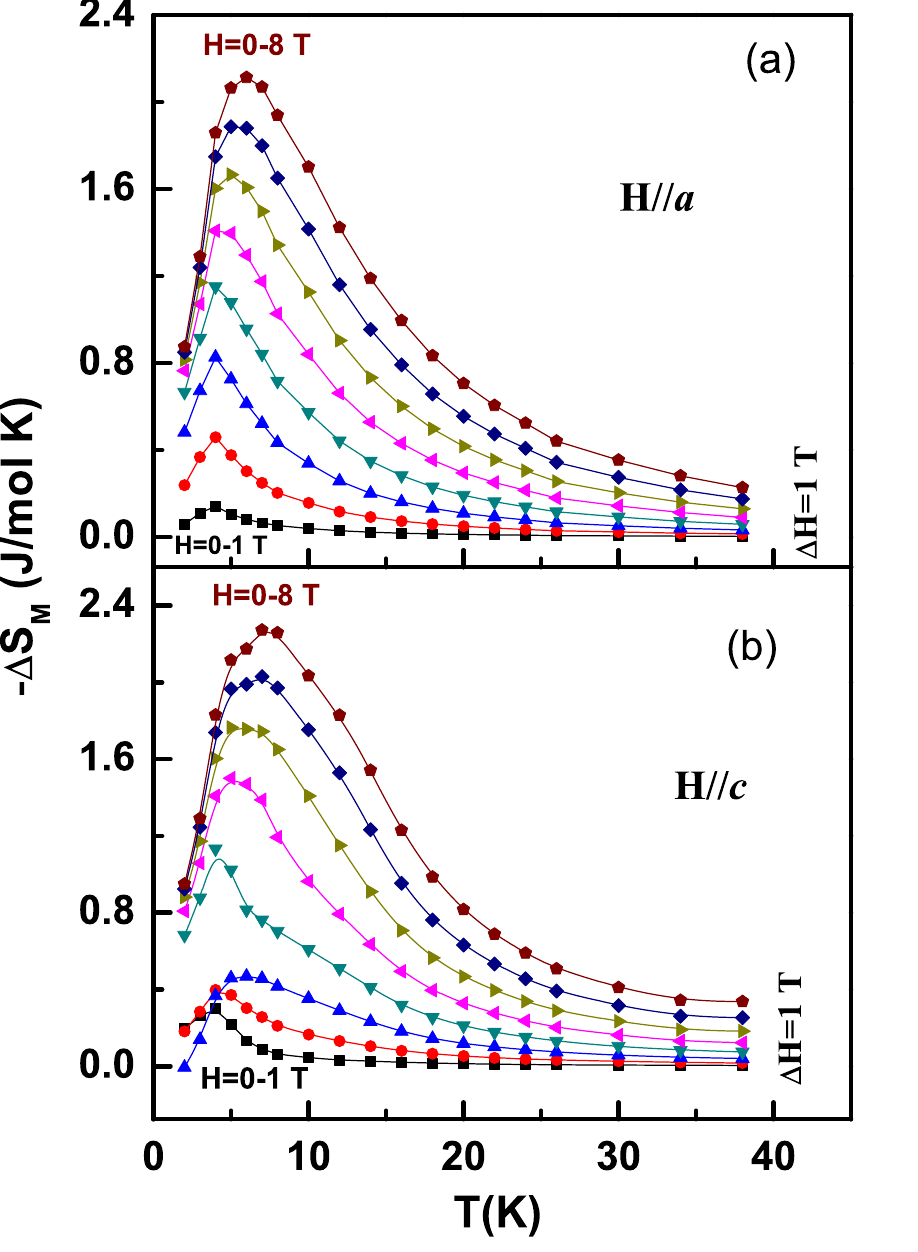}
  \caption{The magnetic entropy change in YbMnO$_3$ for a field change from 0-1 to 0-8 T along the $a$ axis (a) and $c$ axis (b).}\label{Fig.10}
\end{figure}

\begin{figure}[b]
  \includegraphics[width=0.35\textwidth]{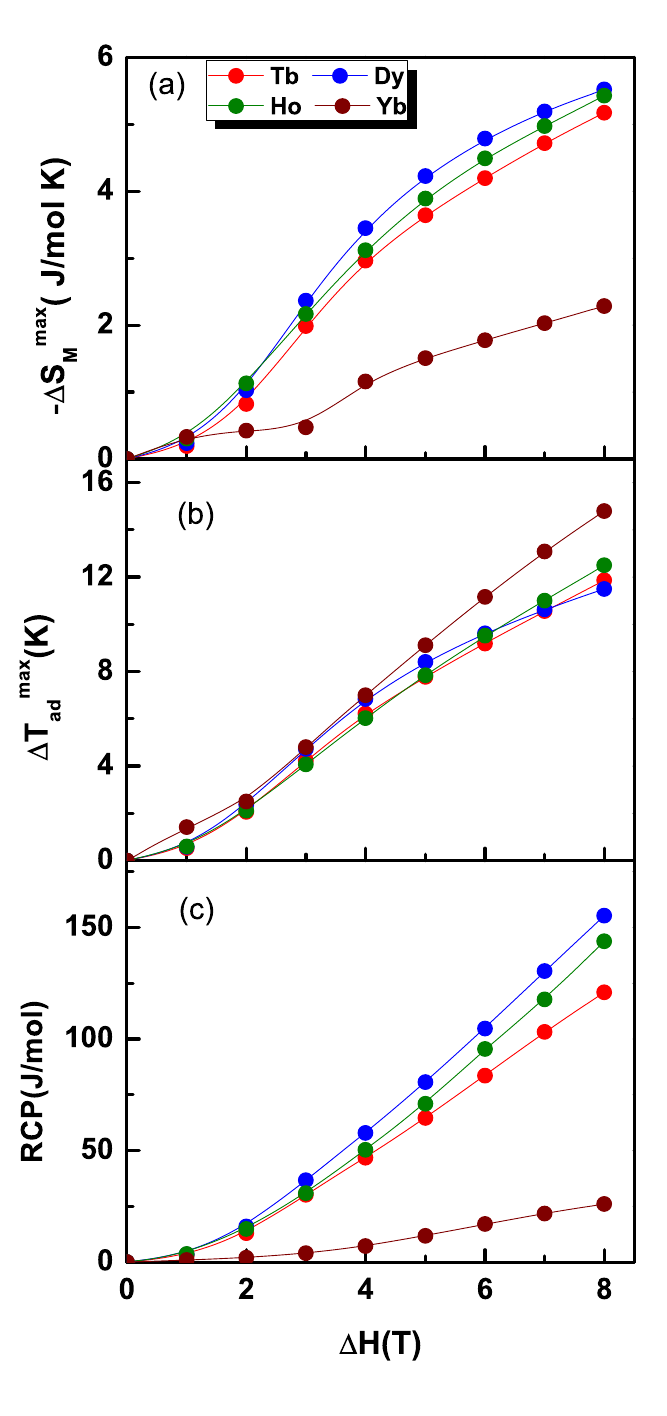}
  \caption{The maximum of magnetic entropy change (a), maximum adiabatic temperature change (b) and relative cooling power (c) as a function of field for $R$MnO$_3$ crystals with $H$ parallel to easy axis.}\label{Fig.11}
\end{figure}
We now briefly discuss the magnetic and magnetocaloric properties of YbMnO$_3$ crystal. Similar to other  multiferroic manganites, magnetic properties of rare-earth sublattice in YbMnO$_3$ is sensitive to the direction of applied field with respect to crystallographic axis.  Figure 8 shows the  field dependence of  magnetization for YbMnO$_3$ at different temperatures. One can see that the nature of magnetic response of the crystal with field along $a$ and $c$ axes are very different.  Unlike HoMnO$_3$, $M$ is slightly larger along the $c$ axis as compared to basal plane and two field-induced transitions are clearly visible.  However, $M$ along $a$ axis increases smoothly with $H$ similar to that in HoMnO$_3$ crystal.  The transition at high-field ($\sim$3 T) is extremely sharp (step-like) below $T_{N}^{Yb}$ ($=$3.6 K), shifts progressively towards higher field with increasing $T$ and  clearly visible up to 14 K [Fig. 8(b)]. It is also interesting to note that  $M$ does not show  saturation-like behavior up to 8 T and the value $M$ at 2 K and 8 T for both the axes are comparable and about 37-39$\%$ of the expected moment.  This behavior is quite different from other multiferroic manganites where $M$ along two axes are significantly different and  the observed high-field value of $M$ for $H$$\Vert$$e$ is close to expected moment. This suggests that YbMnO$_3$ is magnetically less anisotropic and the AFM interaction along both the axes is strong. Indeed from the magnetization measurements in the paramagnetic state, we observe that $\theta_{cw}$ is large and negative along both the crystallographic axes and their values are 195 and 225 K within the basal plane and along $c$ axis, respectively. In order to determine the order of field-induced magnetic  transition, we have also done the Arrott plots of magnetization data of YbMnO$_3$ [Fig. 9]. The nature of the Arrott plots for $H$ along $a$ axis is qualitatively similar to those for HoMnO$_3$ crystal but for $H$ along $c$ axis the low-field peak is quite prominent and the change in slope from positive to negative is abrupt, marked by a much sharper triangular-shaped minimum at $H_c$.  However, unlike HoMnO$_3$, the linear extrapolation of high-field magnetization data to $H$$=$0 reveals a non-zero positive value of $M$ up to 2$T_{N}^{Yb}$ for $H$ along $c$ axis. Figure 10 illustrates the temperature dependence of $\Delta S_{M}$  in a magnetic field change up to 8 T. $\Delta S_{M}$ reaches a negative maximum value $\sim$2.3 J mol$^{-1}$ K$^{-1}$ for $\Delta$H$=$8 T. Though the magnitude of $\Delta S_{M}$ is smaller for YbMnO$_3$ due to the smaller total angular momentum quantum number, the overall nature of $T$ and $H$ dependence of $\Delta S_{M}$ is similar to HoMnO$_3$.  We observe that the value of $\Delta T_{ad}$ is significantly large ($\sim$15 K) in spite of its smaller $\Delta S_{M}$ and RCP (26 J mol$^{-1}$).
In order to compare and contrast the nature of MCE among multiferroic manganites, the variation of $\Delta$S$_{M}^{max}$, $\Delta$T$_{ad}^{max}$ and RCP for $H$$\Vert$$e$ have been plotted in Fig. 11 which shows that these parameters increase monotonically with $\Delta$H. Though the values of $\Delta$S$_{M}^{max}$, $\Delta$T$_{ad}^{max}$  and RCP depend on the system, the nature of their field dependence is qualitatively similar. It is clear from the figures that among the four systems we have studied, DyMnO$_3$, TbMnO$_3$ and HoMnO$_3$ are most suitable as low temperature refrigerant due to the high values of magnetocaloric parameters at a moderate field change.
\section{Conclusions}
In conclusion, the detailed analysis of magnetization data shows that the magnetic interaction within the rare-earth sublattice in multiferroics $R$MnO$_3$ is highly anisotropic (except $R$$=$Yb). Above a critical field $H_c$, $R$MnO$_3$ undergo a field-induced first-order metamagnetic transition from AFM to FM state  and a second-order PM-FM transition  along with huge magnetic entropy change. Depending on $T$  and the direction of applied field, $\Delta S_{M}$ can be negative or positive, i.e.,  $R$MnO$_3$ exhibit both normal and inverse MCE. Except at low temperatures well below $T_{N}^{R}$,  $\Delta S_{M}$ is negative and large in the field-induced FM state. For orthorhombic DyMnO$_3$ and TbMnO$_3$, the inverse MCE is small, decreases with $H$ and eventually crosses over to normal one above $H_c$ for $H$$\Vert$$e$ while it is large and increases with $H$ for $H$$\Vert$$h$. In contrary to this, hexagonal HoMnO$_3$ and YbMnO$_3$ may show small inverse MCE  only below $H_c$ and at very low temperature for $H$$\Vert$$c$.  Both orthorhombic and hexagonal multiferroics, except YbMnO$_3$, exhibit giant MCE, and large adiabatic temperature change and relative cooling power for $H$$\Vert$$e$. The large values of these parameters, negligible hysteresis and highly insulating nature suggest that the multiferroic manganites could be potential materials for magnetic refrigeration in the low-temperature region. The present results also show that orthorhombic and hexagonal multiferroic manganites  can be differentiated based on the magnetic and magnetocaloric properties of rare-earth sublattice.
\section{Acknowledgements}
The authors would like to thank A. Pal  for technical help during sample preparation and measurements.  V. Ganesan would like to thank DST, India for financial assistance for the physical property measurement system facility at UGC-DAE CSR, Indore.

\end{document}